\documentclass[structabstract,longauth]{aa}
\usepackage{epsfig}
\usepackage{graphicx}
\usepackage{txfonts}

\begin{document}

\title{AGILE detection of  {GeV} $\gamma$-ray emission from the SNR W28}

  \author{A.\ Giuliani   \inst{1,*}
    \and M.\ Tavani      \inst{2,3}
    \and A.\ Bulgarelli  \inst{5}
    \and E.\ Striani      \inst{3,12}
    \and S.\ Sabatini      \inst{2}
    \and M.\ Cardillo     \inst{2}
    \and Y. \ Fukui           \inst{7}
    \and A.\  Kawamura   \inst{7}
    \and \\ A.\ Ohama      \inst{7}
    \and N.\ Furukawa  \inst{7}
    \and K.\ Torii           \inst{7}
    \and F. \ A.\ Aharonian \inst{18,19}
    \and F.\ Verrecchia  \inst{15}
     \and  \\ A.\ Argan     \inst{2}
     \and G.\ Barbiellini \inst{6}
     \and P.\ A.\ Caraveo \inst{1}
     \and P.\ W.\ Cattaneo \inst{8}
     \and A.\ W.\ Chen    \inst{1}
     \and V.\ Cocco       \inst{12}
     \and E.\ Costa       \inst{2}
     \and  F.\ D'Ammando   \inst{2,3}
     \and  E.\ Del Monte   \inst{2}
     \and G.\ De Paris    \inst{2}
     \and G.\ Di Cocco    \inst{5}
     \and I.\ Donnarumma  \inst{2}
     \and Y.\ Evangelista \inst{2}
     \and M.\ Feroci      \inst{2}
     \and  M.\ Fiorini     \inst{1}
     \and T.\ Froysland   \inst{3,4}
     \and F.\ Fuschino    \inst{5}
     \and M.\ Galli       \inst{10}
     \and F.\ Gianotti    \inst{5}
     \and C.\ Labanti     \inst{5}
     \and Y.\ Lapshov     \inst{2}
     \and F.\ Lazzarotto  \inst{2}
     \and P.\ Lipari      \inst{11}
     \and F.\ Longo       \inst{6}
     \and M.\ Marisaldi   \inst{5}
     \and S.\ Mereghetti  \inst{1}
     \and A.\ Morselli    \inst{12}
     \and E.\ Moretti    \inst{6}
     \and L.\ Pacciani    \inst{2}
     \and A.\ Pellizzoni  \inst{9}
     \and F.\ Perotti     \inst{1}
     \and P.\ Picozza     \inst{3,12}
     \and M.\ Pilia       \inst{9}
     \and M.\ Prest       \inst{13}
     \and G.\ Pucella     \inst{14}
     \and M.\ Rapisarda   \inst{14}
     \and A.\ Rappoldi    \inst{8}
     \and P.\ Soffitta    \inst{2}
     \and M.\ Trifoglio   \inst{5}
     \and A.\ Trois       \inst{2}
     \and  E.\ Vallazza    \inst{6}
     \and S.\ Vercellone  \inst{17}
     \and  V.\ Vittorini   \inst{2,4}
     \and A.\ Zambra      \inst{1}
     \and D.\ Zanello     \inst{9}
     \and C.\ Pittori     \inst{15}
     \and P.\ Santolamazza \inst{15}
     \and P.\ Giommi      \inst{15}
     \and S.\ Colafrancesco \inst{15}
     \and L.\ Salotti     \inst{16}
    }

    \institute{INAF/IASF--Milano, Via E.\ Bassini 15, I-20133 Milano, Italy
   \and INAF/IASF--Roma, Via Fosso del Cavaliere 100, I-00133 Roma, Italy
   \and Dip. di Fisica, Univ. ``Tor Vergata'', Via della Ricerca
  Scientifica 1, I-00133 Roma, Italy
   \and CIFS--Torino, Viale Settimio Severo 3, I-10133, Torino, Italy
   \and INAF/IASF--Bologna, Via Gobetti 101, I-40129 Bologna, Italy
   \and Dip. di Fisica and INFN, Via Valerio 2, I-34127 Trieste, Italy
   \and Department of Astrophysics, Nagoya University, Chikusa-ku, Nagoya 464-8602, Japan
   \and INFN--Pavia, Via Bassi 6, I-27100 Pavia, Italy       %
   \and INAF-Osservatorio Astronomico di Cagliari, localita' Poggio dei Pini, strada 54, I-09012 Capoterra, Italy
   \and ENEA--Bologna, Via dei Martiri di Monte Sole 4, I-40129 Bologna, Italy
   \and INFN--Roma ``La Sapienza'', Piazzale A. Moro 2, I-00185 Roma, Italy
   \and INFN--Roma ``Tor Vergata'', Via della Ricerca Scientifica 1, I-00133 Roma, Italy
   \and Dip. di Fisica, Univ. dell'Insubria, Via Valleggio 11, I-22100 Como, Italy
   \and ENEA--Roma, Via E. Fermi 45, I-00044 Frascati (Roma), Italy
   \and ASI--ASDC, Via G. Galilei, I-00044 Frascati (Roma), Italy
   \and ASI, Viale Liegi 26 , I-00198 Roma, Italy
   \and INAF/IASF--Palermo, Via La Malfa 153, I-90146 Palermo, Italy
   \and Max-Planck-Institut f\"ur Kernphysik, 69117 Heidelberg, Germany.
   \and Dublin Institute for Advanced Sudies, Dublin 2, Ireland
}

\offprints{giuliani@iasf-milano.inaf.it, *Corresponding Author: Andrea Giuliani}
\date{Received : 13 February 2010, Accepted : 5 may 2010}
\abstract{}
{Supernova remnants (SNRs) are believed to be the main sources of Galactic cosmic rays.
Molecular clouds associated with SNRs can produce gamma-ray emission through the interaction of accelerated particles with the concentrated gas.
The middle aged SNR W28, for its associated system of dense molecular clouds, provides an excellent opportunity  to test this hypothesis.}%
{We present the AGILE/GRID observations of SNR W28,  and  compare them with observations at other wavelengths (TeV and $^{12}$CO $(J=1\rightarrow0)$ molecular line  emission).
}%
{ 
The gamma-ray flux detected by AGILE from the dominant source associated with W28 is  (14 $\pm$ 5) $\times$ 10$^{-8}$ ph cm$^{-2}$ s$^{-1}$ for $E > 400$ MeV.
This source is positionally well correlated with the TeV emission observed by the HESS  telescope.
The local variations of the GeV to TeV flux ratio suggest a difference between the CR spectra of
the north-west and south molecular cloud complexes. %
A model based on a hadronic-induced interaction and diffusion with two molecular clouds at different distances from the W28 shell can explain both the morphological and spectral features observed by AGILE in the MeV-GeV energy range and by the HESS telescope in the TeV energy range.
{The combined set of AGILE and H.E.S.S. data strongly support a hadronic
       model for the gamma-ray production in W28.}}
{}%
\keywords{gamma rays: observations - Supernova Remnants: individual: W28}
\authorrunning{A. Giuliani et al.}
\titlerunning{High energy emission of SNR W28.}
\maketitle

\section{Introduction}

\begin{figure}[ht!]
\begin{center}
\includegraphics[width=9cm]{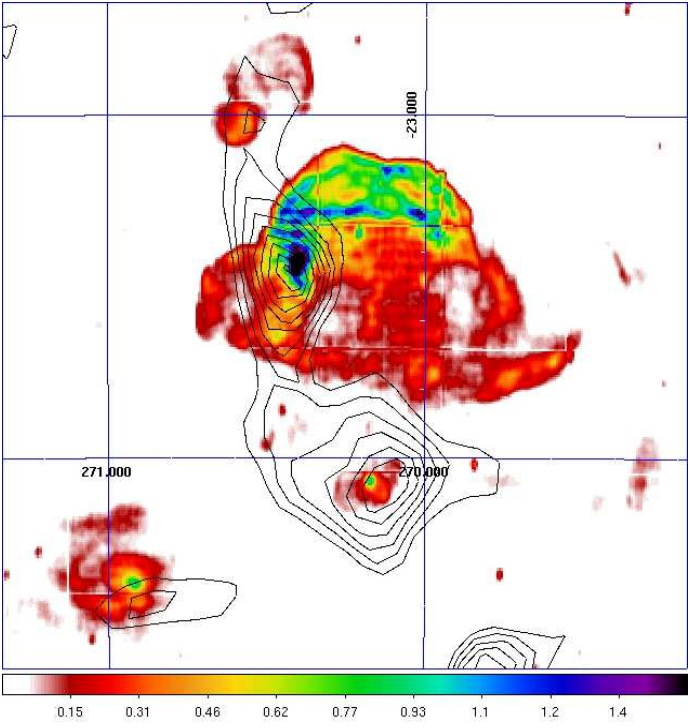}
\caption{Map of VLA 90 cm radio emission from SNR W28  { in celestial J2000 coordinates R.A. and {Dec}. (Colours indicate the intensity in Jy beam$^{-1}$)}. 
The black contours show the CO intensity emission obtained with the NANTEN radio telescope which trace molecular
clouds, integrated over the velocity interval of 3-27 km s$^{-1}$ (contours vary between 90 and 170 K deg with a step size of 10 K deg). 
Two concentrations of CO emission are clearly visible near the positions $\alpha, \delta$ = 270.4, -23.4 (cloud N) and  $\alpha, \delta$ = 270.2, -24.1  (cloud S).
}
\end{center}
\label{radio}
\end{figure}

SNR W28, also known as G6.4-0.1, is a middle aged supernova remnant (with an age of at least 35 thousand years), located at  a distance between 1.8 and 3.3 kpc in the inner region of the Galaxy (\textit{l,b} = 6.71, -0.05).
W28 is a mixed-morphology SNR, with large (about 48 arcmin) apparent dimensions (see \cite{green09}).
At radio wavelength a shell structure, where the interaction of the SNR ejecta and the ISM creates a shock, is clearly visible (see Figure \ref{radio}).
The spectral index between 328 and 1415 MHz is $\alpha=-0.35 \pm 0.18$ and shows large local variations correlated with the flux density (\cite{dubner00} ).
The $^{12}$CO $(J=1\rightarrow0)$ molecular line observations taken by the NANTEN telescope reveals a system of molecular clouds  associated with the SNR (\cite{mizuno04}).
Contours in Figure \ref{radio} show the intensity of the $^{12}$CO line emission integrated over the velocity range 3 to 27 km s$^{-1}$ corresponding to the dynamical distances compatible with W28.
The emission appears concentrated in two main complexes near the north-west (cloud N) and south (cloud S) part of the SNR.
Evidence of interaction between the remnant and the system of molecular clouds is given  {by the detection of 1720 MHz OH {maser} emission (\cite{frail94})}
and by the observation of an  {unusually} high value of the ratio CO $ {(J=3-2)/(J =2-1)}$ (\cite{arikawa99}).
\\Two TeV sources have been detected by the HESS Cherenkov telescope near W28 (\cite{aharonian06}).
The sources HESS J1801-233  and HESS J1800-240 (A, B and C) are very well correlated with, respectively, the N and S molecular clouds  resolved by the NANTEN telescope (\cite{aharonian08}).
In the MeV-GeV energy range a source compatible with W28 was reported in the Third EGRET catalog, 3EG J1800-2338 (\cite{hartman99}).
The flux measured by EGRET above 100 MeV was $61.3 \pm 6.7$ ph cm$^{-2}$ s$^{-1}$, with a photon spectral index of 2.10.
In both the Fermi LAT Bright Source List (0FGL J1801.6-2327, see \cite{fermi_catalog}) and the First AGILE Catalog (1AGL J1803-2258, see \cite{agile_catalog}) a gamma-ray source centered on the position of cloud N is reported.
\\We present in this Letter the results of deep AGILE observation of  1AGL J1803-2258 which, at energies greater than 400 MeV, has a shape that turns out to be incompatible with a single point source, and is remarkably correlated with both the molecular cloud system seen by NANTEN and the TeV sources detected by HESS.
\begin{figure}[ht!]
\begin{center}
\includegraphics[width=9cm]{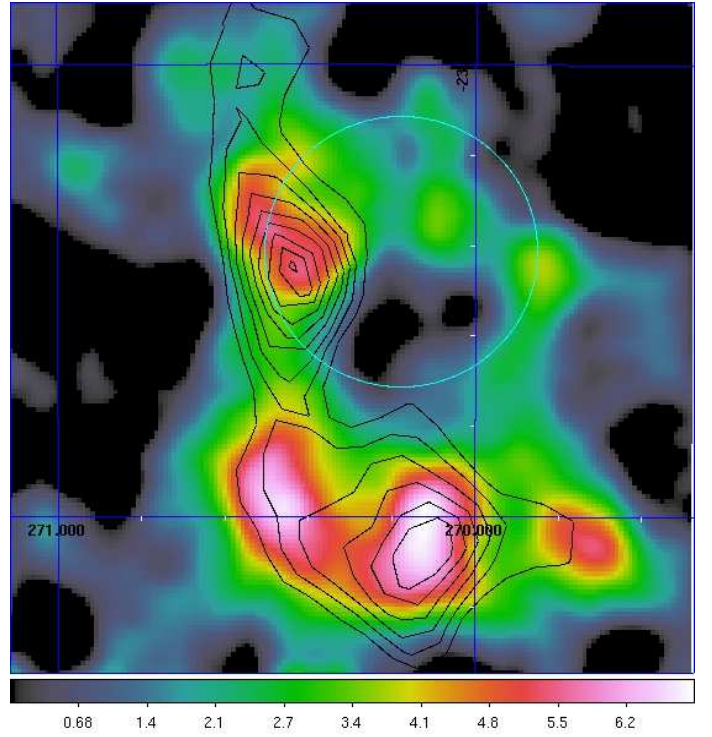}
\end{center}
\caption{Sky map   {in celestial J2000 coordinates, R.A. and {Dec}.,} of the significance   {($\sigma$)} of the TeV  emission detected by the HESS telescope (\cite{aharonian08}).
The blue circle indicates the location of the supernova remnant W28. As in Figure \ref{radio} the black contours show the CO
intensity emission. } \label{tev}
\end{figure}
\section{Data Analysis}

The Gamma-Ray Image Detector (GRID) on board  AGILE satellite (\cite{tavani}) extensively observed  the Galactic plane during the years 2007-2009 at energies greater than 100 MeV with an angular resolution better than the previous gamma-ray telescopes.
Level-1 AGILE-GRID data were analyzed using the AGILE Standard Analysis Pipeline.
The Anticoincidence System (ACS) and a set of hardware triggers perform the first reduction of the high rate of background events (charged particles and albedo gamma-rays) interacting with the instrument.
A dedicated software processes the GRID and ACS signals and performs a reconstruction and selection of events in order to  discriminate between background events and gamma-rays, deriving for the latter the energy and direction of the incoming photons.
A simplified version of this software operates directly on board AGILE in order to reduce the telemetry throughput
(\cite{giuliani06}).
A more complex version of the photon reconstruction and selection software is applied on the ground, producing a photon list containing arrival time, energy and direction of every gamma-ray and the corresponding exposure history.
Counts, exposure and Galactic background gamma-ray maps, the latter based on the Galactic diffuse emission model developed for AGILE (\cite{giuliani04}), were created with a bin-size of 0.$^{\circ}$05 $\times$ 0.$^{\circ}$05 for photons with energy greater than 100 MeV and 400 MeV.
We selected only events flagged as confirmed gamma-ray events. All events collected during the South Atlantic Anomaly or whose reconstructed directions form angles with the satellite-Earth vector smaller than 90$^{\circ}$ are rejected.
In order to derive the average flux and spectrum of the source we ran the AGILE point source analysis software 
 {ALIKE}  {(}\cite{chen10} {)}, based on the maximum likelihood technique as described in  {Mattox et al. (1993)}, over  the whole observing period. 
 {All the flux errors found by ALIKE (and reported in the following) take into account only the statistical uncertainities. We estimate that the systematic errors are on the level of $10 \%$ of the reported fluxes.}

\subsection{Results}

\begin{figure}[ht!]
\begin{center}
\includegraphics[width=9cm]{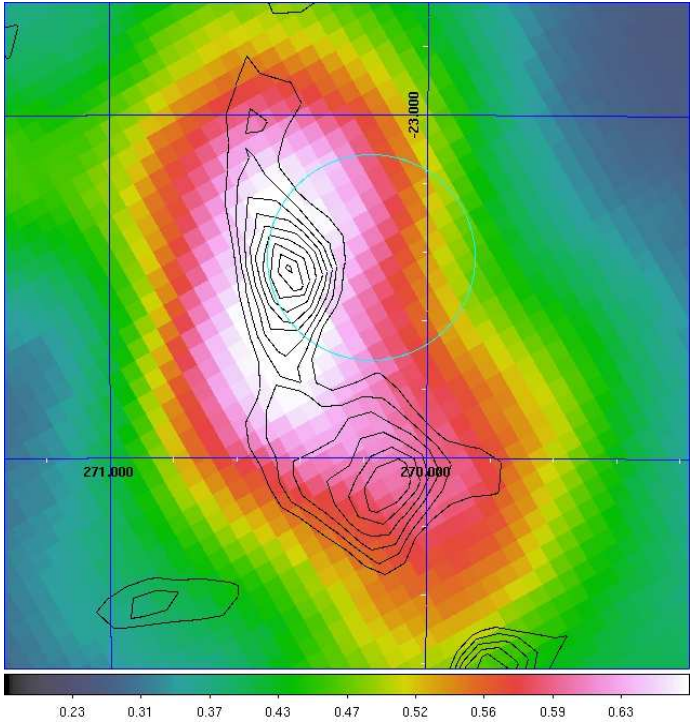}
\end{center}
\caption{Gaussian-smoothed AGILE counts map in Galactic coordinates for the W28 region (counts bin$^{-1}$).
Only photons with energy greater than 400 MeV have been used.
The blu circle indicates the radio location of the supernova remnant W28.
As in Figure  \ref{radio} the black contours give the CO intensity emission.
}
\label{mappa}
\end{figure}

A prominent gamma-ray source associable with W28 (1AGL J1803-2258) is clearly detected
at a significance level of 7 $\sigma$  with an average flux of (40 $\pm$ 11) $\times$ 10$^{-8}$ ph cm$^{-2}$ s$^{-1}$ for E $>$ 100 MeV  {above the prediction of the AGILE diffuse emission model} .
The positional analysis gives an elliptical error-box centered in \textit{l} =$6.0^{\circ}$ \textit{b} =$-0.2^{\circ}$ with a mean radius of about 0.1$^{\circ}$
consistent with both the EGRET and Fermi sources detected in the same region.
\\Figure \ref{mappa} shows the  {counts} map for the source 1AGL J1803-2258 for energies greater than 400 MeV.
At these energies the good angular resolution of the GRID detector allows to perform a morphological analysis of the source, which shows a remarkable correspondence with the TeV emission observed in the same region.
The total flux of the source at energy greater than 400 MeV is (14 $\pm$ 5) $\times$ 10$^{-8}$ ph cm$^{-2}$ s$^{-1}$.
Most of the emission is concentrated in a region coincident with the molecular cloud N (and the TeV source HESS J1801-233).
A weaker gamma-ray diffuse emission appears to be superposed on the molecular cloud S (and the TeV source complex HESS J1800-240).
Performing a likelihood analysis with two point sources fixed at the locations of source N and source S, we find flux  values
$F_N = (10 \pm 3)$ $\times$ 10$^{-8}$ ph cm$^{-2}$ s$^{-1}$ and $F_S = (4 \pm 2)$ $\times$ 10$^{-8}$ ph cm$^{-2}$ s$^{-1}$ above 400 MeV.
The two regions are not clearly resolved in the 100-400 MeV energy range.
Nevertheless, we performed a likelihood analysis assuming the same two point sources as  in the analysis for the $E>400$ MeV energy range.
This yields a flux of (30 $\pm$ 6)  $\times$ 10$^{-8}$ ph cm$^{-2}$ s$^{-1}$ for cloud N and  {a $2 \sigma$} upper limit of 10 $\times$ 10$^{-8}$ ph cm$^{-2}$ s$^{-1}$ for cloud S.

\vskip -2cm
\section{Discussion}
\begin{figure}[ht!]
\begin{center}
\includegraphics[width=9.0cm, height=11cm]{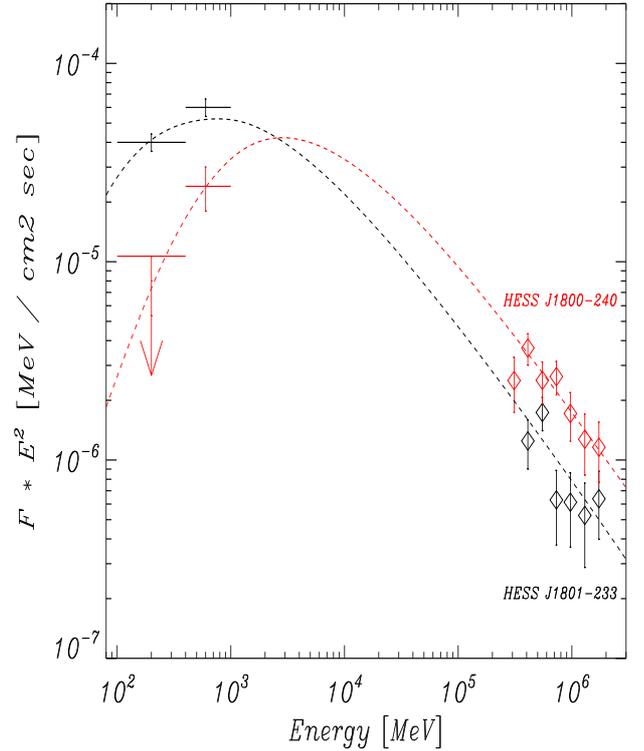}
\end{center}\caption{ {Gamma-ray photon spectrum for W28 (AGILE points: ${1} \sigma$ statistical errors, HESS points are from Aharonian et al. 2008). 
The curves represent the gamma-ray spectra estimated, accordling to the model presented in the text,
for the cloud N (black) and the cloud S (red) assumed at a distance of  4 and 9 pc, respectively,  from the particle acceleration site.}}
\label{spec}
\end{figure}

Figure \ref{spec} shows the spectra for cloud N (black) and cloud S (red) in the GRID and in the HESS energy bands.
While in the TeV energy band the complex of the source HESS 1800-240 is a factor of two brighter than the source HESS 1801-233, in the GeV band the reverse is true.
Within a hadronic scenario,  {and assuming CRs are accelerated by W28 and then escaping,}
the difference between the TeV to GeV flux ratios determined at the locations N and S
implies different proton spectra in the two clouds.
This is not surprising because, for a turbulent medium, the diffusion coefficient of charged  particles is expected to depend on the particle kinetic energy, $D=D(E)$.
As a consequence, in  {the} presence of recent acceleration (less than $10^5$ years),
the proton spectrum  depends strongly on the distance from the accelerator itself (\cite{aharonian96},
\cite{gabici09}).
Hence the difference in the gamma-ray spectra of cloud N and S can be explained by assuming different
distances between the clouds and the SNR.
 {In the following, we  show that this model can explain the
GeV/TeV spectra of the clouds assuming a set of parameters (shown
in Table 1) well compatible with other observational constraints.}
We estimated (using the mathematical formalism developed in
\cite{aharonian96}) the evolution of the protons and nuclei
spectra diffusing in the interstellar medium at different
distances from the supernova remnant.
We assume  {a distance of 2 kpc} and an age of the SNR of 45000 years and that the particles have been injected continuously during the first 10000 years of its life, with a power-law energy spectrum of index  $2.2$.
After the injection we assume that they propagate in the interstellar medium with diffusion coefficient given by :
\begin{equation}
D=D_0  \left(  \frac{E}{10 \; GeV} \right)  ^{\delta}
\end{equation}
with $\delta = 0.5$ and  {a rather slow diffusion coefficient $D_0=10^{26}$ cm$^2$ s$^{-1}$
as indicated by Aharonian and Atoyan (1996) for a dense medium such as that of W28}.
\\In Figure \ref{spec} we compare the  results of this model with the AGILE  and HESS spectral data. The dashed lines correspond to a cloud accelerator distance of 4 (in black) and 9 (in red) parsecs.  {Note that a different set of parameters, the cloud-distances and the SNR-age, can lead to similar results (in general, any set of these parameters with the same ratio distances/$\sqrt{SNR \; age}$ can reproduce the curves shown in Fig \ref{spec}). 
We choose here the distance of the cloud S as the minimum distance compatible with the observations. 
A distance of 4 pc for cloud N (which requires assuming a projection effect of a few parsecs) and an age of 45,000 years for W28 are then required in order to reproduce the GeV to TeV ratio of the clouds.} 
We set the mass of the cloud S to be larger (by a factor 2.25) than that
of cloud N,
 {as can be inferred from the TeV fluxes and from the {proton} density predicted by this model at 4 and 9 pc from the SNR shell}.
The distances used in this model are consistent with the position of the clouds as observed in the radio maps (assuming an Earth-SNR distance of 2 kpc, 10 parsecs corresponds to 0.29 deg),
taking into account posible projection effects.
Assuming a mass for the two molecular clouds,  it is possible to evaluate the total energy carried by the protons accelerated by W28.
We used $2.0 \times 10^4 M_{\sun}$  for cloud N and  $4.5 \times 10^4 M_{\sun}$ for cloud S, in agreement with the estimate based on NANTEN data given in \cite{aharonian08}.
This leads to a proton total energy of $3.3 \times 10^{49}$ erg,
a value in good agreement with the CR energy release from a SNR expected by theoretical considerations.
\\ {Finally, we  briefly consider here a leptonic scenario for the gamma-ray emission from W28.
Relativistic electrons can produce gamma rays through inverse Compton (IC) and/or Bremsstrahlung in a dense target. 
The cooling times for these processes  are :
\begin{equation}
t_{br}=4 \times 10^7   \; \left(   \frac{1 \; \mathrm{cm}^{-3}}{n}  \right)   \;  \mathrm{yr}
\end{equation}
\begin{equation}
t_{IC}(E)=3 \times 10^8  \; \left(   \frac{1 \;\mathrm{eV}  \;
\mathrm{cm}^{-3}}{w_r}  \right) \left(  \frac{1 \; \mathrm{GeV}}{E} \right)   \; \mathrm{yr}
\end{equation}
where $n$ is the gas density and $w_r$ the energy density of the radiation field {(note that eq. 3 refers to IC scattering in the Thompson regime)}. 
For comparison the cooling time for protons producing gamma-rays by proton-proton interactions is  :
\begin{equation}
t_{pp}=1.6 \times 10^8   \; \left(   \frac{1 \; \mathrm{cm}^{-3}}{n}  \right)    \;  \mathrm{yr}
\end{equation}
Inverse Compton is disfavoured because%
the strong detected correlation (especially at TeV energies) between the gamma ray intensity and dense gas
{is not expected for evolved SNR like W28}.
Bremsstrahlung can, in principle, produce the observed gamma-ray emission. 
Assuming that TeV gamma-rays are produced by $\sim$ 10 TeV electrons, Bremsstrahlung dominates over the IC in very dense regions ($n>1000$ $\rm cm^{-3}$). 
However such energetic electrons are expected to be produced only at the early epochs of the SNR {(\cite{gabici09})}, and they can only marginally survive for the age of this source due to strong synchrotron and IC cooling.
Furthermore, in order to explain the observed  fluxes, a leptonic model would require a very large e/p ratio in accelerated particles (larger than 0.25, comparing eq. 2 and 4) while a ratio near 0.01 is expected from the CR measurements. 
}
\vskip -0.3cm
\begin{table}[ht!]
\caption{Parameters of the model used to fit the AGILE and HESS spectral data, see fig \ref{spec}.}
\begin{tabular}{ll}
\hline  &  \\
SNR age & 45000 yr \\
 {SNR distance} &   {2}~kpc \\
CR total energy   &  $3.3 \times 10^{49}$~erg\\
CR Injection index  & 2.2  \\
Diffusion coeff.  @ 10 GeV  ($D_0$)  & $10^{26}$  $\rm cm^2 s^{-1}$\\
Diffusion coeff.  spectral index  ($\delta$) & 0.5  \\
Cloud N (mass, distance  {from the W28 shell})  & $2.0 \times 10^ {4} M_{\sun}$, $4$ pc \\
Cloud S (mass, distance  {from the W28 shell})  &  $4.5 \times 10^ {4} M_{\sun}$, $9$ pc \\
 &  \\
\hline
\end{tabular}
\end{table}

\section{Conclusions}
The AGILE deep observations of SNR W28 determine the existence of several contributions to the
gamma-ray emission from this SNR.
A highly significant gamma-ray source (7 $\sigma$ for E$>$ 100 MeV) is associated with W28.
This source has an average  flux of (40 $\pm$ 11) $\times$ 10$^{-8}$  ph cm$^{-2}$ s$^{-1}$ during the period covered by the AGILE observations.
AGILE also detected, for photon energies above 400 MeV,
a SE extension corresponding to a massive molecular cloud.
We proposed a model based on a hadronic-induced interaction with two molecular clouds
adjacent to the SNR in order to fit the observations.
This model explains the morphological and spectral features detected both by AGILE i
n the MeV-GeV energy range and by the HESS telescope in the TeV energy range.
Setting the distances and masses of the two main molecular clouds to values compatible
with the radio CO observations we can estimate that the total energy in protons is
$3.3 \times 10^{49}$ erg.
 {Our data and analysis 
provides strong support to a 
hadronic origin of the gamma-ray emission from W28.}

\begin{acknowledgements}
The AGILE Mission is funded by the Italian Space Agency (ASI) with
scientific and programmatic participation by the Italian Institute
of Astrophysics (INAF) and the Italian Institute of Nuclear
Physics (INFN).Investigation carried out with partial support
by the ASI contract no. I/089//06/2.
\end{acknowledgements}

\bibliography{matt}

\begin{thebibliography}{}

\bibitem[{Abdo} et~al., 2009]{fermi_catalog}
{Abdo}, A.~A. et al.
(2009).
\newblock {\em \apjs}, 183, 46--66.

\bibitem[{Aharonian} et~al., 2008]{aharonian08}
{Aharonian}, F. et al.
  (2008).
\newblock {\em \aap}, 481, 401--410.

\bibitem[{Aharonian} et~al., 2006]{aharonian06}
{Aharonian}, F. et al. (2006).
\newblock {\em \apj}, 636, 777--797.

\bibitem[{Aharonian} \& {Atoyan}, 1996]{aharonian96}
{Aharonian}, F.~A. and {Atoyan}, A.~M. (1996).
\newblock {\em \aap}, 309, 917--928.

\bibitem[{Arikawa} et~al., 1999]{arikawa99}
{Arikawa}, Y. et al. (1999).
\newblock {\em \pasj}, 51, L7--L10.

\bibitem[{Chen} et~al., 2010]{chen10}
{Chen}, A. et al. (2010).
\newblock {In preparation}.

\bibitem[{Dubner} et~al., 2000]{dubner00}
{Dubner}, G.~M. et al.  (2000).
\newblock {\em \aj}, 120, 1933--1945.

\bibitem[{Frail} et~al., 1994]{frail94}
{Frail}, D.~A., {Goss}, W.~M., and {Slysh}, V.~I. (1994).
\newblock {\em \apjl}, 424, L111--L113.

\bibitem[{Gabici} et~al., 2009]{gabici09}
{Gabici}, S., {Aharonian}, F.~A., and {Casanova}, S. (2009).
\newblock {\em \mnras}, 396, 1629--1639.

\bibitem[{Giuliani} et~al., 2004]{giuliani04}
{Giuliani}, A. et al. (2004).
\newblock {\em Mem.  S.A.It. Sup.}, 5,
  135--+.

\bibitem[{Giuliani} et~al., 2006]{giuliani06}
{Giuliani}, A. et al.  (2006).
\newblock {\em Nucl. Instr. and Meth. A}, 568,
  692--699.

\bibitem[Green 2009]{green09} Green, D.~A.\ 2009, Bulletin of 
the Astronomical Society of India, 37, 45 

\bibitem[{Hartman} et~al., 1999]{hartman99}
{Hartman}, R.~C. et al. (1999).
\newblock {\em \apjs}, 123, 79--202.

\bibitem[{Mattox} et~al., 1993]{mattox93}
{Mattox}, J.~R. et al. (1993).
\newblock {\em \apj}, 410, 609--614.

\bibitem[{Mizuno} \& {Fukui}, 2004]{mizuno04}
{Mizuno}, A. and {Fukui}, Y. (2004).
\newblock {\em ASP Conference
  Series}, 317, 59-+.

\bibitem[{Pittori} et~al., 2009]{agile_catalog}
{Pittori}, C. et al. (2009).
\newblock {\em \aap}, 506, 1563--1574.

\bibitem[{Tavani} et~al., 2008]{tavani}
{Tavani}, M. et al. (2008).
\newblock {\em Nucl. Instr. and Meth.  A}, 588,
  52--62.

\end{thebibliography}
\bibliographystyle{and}

\end{document}